# Psychiatric Illnesses as Disorders of Network Dynamics

## Running head: Mental Illnesses as Disorders of Network Dynamics

**Daniel Durstewitz[1,2*], Quentin J.M. Huys[3,4], Georgia Koppe[1,5*]**

[1]Department of Theoretical Neuroscience, Central Institute of Mental Health, Medical Faculty Mannheim, Heidelberg University, Germany

[2]Faculty of Physics and Astronomy, Heidelberg University, Germany

[3]Centre for Addiction Disorders, Department of Psychiatry, Psychotherapy and Psychosomatics, Hospital of Psychiatry, University of Zurich

[4]Translational Neuromodeling Unit, Institute for Biomedical Engineering, University of Zurich and ETH Zurich

[5]Department of Psychiatry and Psychotherapy, Central Institute of Mental Health, Medical Faculty Mannheim, Heidelberg University, Germany

[*]corresponding authors: daniel.durstewitz@zi-mannheim.de, georgia.koppe@zi-mannheim.de

## Abstract

This review provides a dynamical systems perspective on psychiatric symptoms and disease, and discusses its potential implications for diagnosis, prognosis, and treatment. After a brief introduction into the theory of dynamical systems, we will focus on the idea that cognitive and emotional functions are implemented in terms of dynamical systems phenomena in the brain, a common assumption in theoretical and computational neuroscience. Specific computational models, anchored in biophysics, for generating different types of network dynamics, and with a relation to psychiatric symptoms, will be briefly reviewed, as well as methodological approaches for reconstructing the system dynamics from observed time series (like fMRI or EEG recordings). We then attempt to outline how psychiatric phenomena, associated with schizophrenia, depression, PTSD, ADHD, phantom pain, and others, could be understood in dynamical systems terms. Most importantly, we will try to convey that the dynamical systems level may provide a central, hub-like level of convergence which unifies and links multiple biophysical and behavioral phenomena, in the sense that diverse biophysical changes can give rise to the same dynamical phenomena and, vice versa, similar changes in dynamics may yield different behavioral symptoms depending on the brain area where these changes manifest. If this assessment is correct, it may have profound implications for the diagnosis, prognosis, and treatment of psychiatric conditions, as it puts the focus on dynamics. We therefore argue that consideration of dynamics should play an important role in the choice and target of interventions.

# 1. Introduction

Time is a key feature in physiological, neural and behavioral processes in general, and of psychiatric illnesses in particular. First, neural and behavioral activity always evolves in time, at temporal scales reaching from milliseconds to days or months. Understanding the principles governing this temporal evolution is crucial to understanding physiological mechanisms as well as computational and cognitive processes, and for predicting future states of the system. Second, psychiatric symptoms become signs of burdensome disease only if they persist long enough, and hence the time periods for which symptoms have been present are critical to the diagnostic process. Third, the burden that arises from psychiatric diseases is to a large extent due to their chronicity. Part of the chronicity is expressed through a relapsing-remitting course over time, where symptoms come and go for decades, repeatedly torpedoing individuals' attempts at building their life's projects. Fourth, understanding the individual developmental trajectories that lead into psychiatric disease, as well as the shorter-scale trajectories along which psychiatric symptoms dis- and reappear, is likely to be a key in prevention and identification of causal factors. Hence, critical aspects of mental health relate to dynamics. Yet, the consideration of dynamics in these settings barely exists.

Dynamical systems theory (DST) is a mathematical framework that deals with systems which evolve in time, be they at subcellular, neural, behavioral or societal levels. Such systems can be described by a set of differential (if formulated in continuous time) or difference (if in discrete time) equations. DST provides a powerful and general mathematical language and toolbox for examining phenomena in such systems which are generic, that is, largely independent from their specific realization in the physical world. These phenomena include, for instance, oscillations, synchronization among units of a system, attractor states, phase transitions, or deterministic chaos. Although generic and formulated in an abstract language, these phenomena are not merely conceptual or even metaphorical, but 'real' in the sense that they can be *measured* and *inferred* from observed data, and used to predict future developments. They are experimentally and clinically accessible and quantifiable processes.

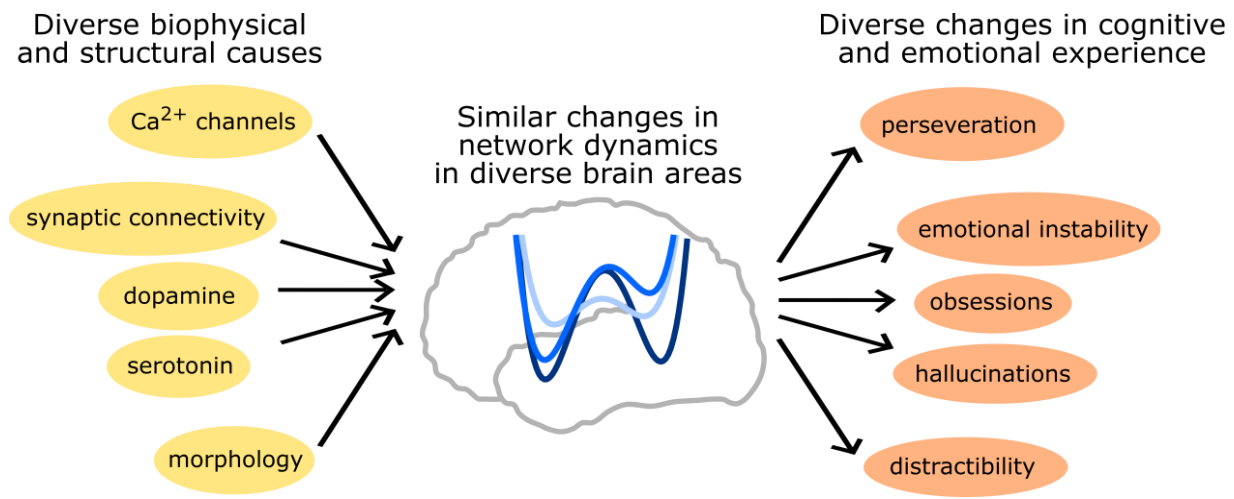

**Fig. 1. Network dynamics as a layer of convergence.** A number of different physiological and structural processes (left) may give rise to similar alterations in network dynamics (center) which, depending on where in the brain they manifest, may give rise to a variety of different cognitive and emotional processes and psychiatric symptoms.

One central tenet in theoretical neuroscience is that computational and cognitive properties of the brain are implemented in terms of the underlying network dynamics – in a sense, DST constitutes the 'computational language' of the brain, a universal model of computation, just like the Turing Machine framework on which most of traditional computer science rests. As such, DST provides a language within which psychological and mental entities and processes can be expressed and, potentially, understood in their mechanisms. Moreover, as we will try to argue in this article, DST may serve as a kind of hub, a central layer of convergence or level of nervous system description, at which psychiatric symptoms and illnesses could be explained, understood, classified, predicted, and treated. A layer of convergence in the sense that 1) a number of very different, seemingly unrelated physiological and anatomical processes may give rise to similar alterations in network dynamics (Fig. 1). This may explain why quite different causal factors and pathogenic routes may give rise to similar symptoms. At the same time, 2) the same changes in network dynamics may be associated with a variety of quite different symptoms (Fig. 1), depending on the brain areas most affected by these alterations, and the type of information processed by the underlying networks. For instance, while hyper-stable attractor states in auditory areas may cause tinnitus, the same alterations in orbitofrontal cortex may be associated with perseveration of suboptimal responses. The large degree of comorbidity found among different psychiatric domains may potentially also be explained this way: Perhaps the brain becomes generally vulnerable to a specific type of alteration of its dynamical regimes (e.g., due to a transmitter imbalance) – depending on which brain areas are affected most by these dynamical alterations, they will find their incarnation in different bundles of symptoms which may change over time.

This emphasis on the *dynamical systems level* also bears important implications for the treatment of mental illness. If we come to view mental illnesses (partly or largely) as *disorders of network dynamics*, then alteration of dynamics should be a central target of therapeutic intervention (although, of course, the channels of intervention would remain neurobiological and/or behavioral). Because dynamics can be altered in similar ways through potentially counter-intuitively different routes, this may open up new paths for intervention, and may also explain why quite different treatments could have similar effects. For instance, if the goal were to restore functional network dynamics which were initially deterred by alterations in D2 receptor affinity, pharmacologically altering Na$^+$ currents might serve just as well to re-establish the normal functional regime as would reversing the original D2 receptor deficit (i.e., these quite different biophysical routes may be functionally equivalent in terms of network dynamics). DST excels in dealing with such counter-intuitive, hard-to-predict phenomena that are characteristic of complex systems consisting of many interacting feedback loops, such as the brain.

This article is structured as follows: We will first give a general overview over formal DST concepts and their visualization, mostly from a neuroscience perspective. Whilst introducing various dynamical phenomena in more detail in the third section, we will then discuss how such systems can represent and implement cognitive and neuro-computational functions, and how a variety of psychiatric symptoms could potentially be explained in terms of alterations in network dynamics. Thus, while the focus will be on dynamics at the neural level where it can build a direct bridge from biophysical properties to mental phenomena, we will also highlight how DST could be applied more generally to explain (and predict) phenomena directly at the behavioral (or even societal) level without direct reference to a physical substrate (Borsboom, Cramer, Schmittmann, Epskamp, & Waldorp, 2011; Nelson, McGorry, Wichers, Wigman, & Hartmann, 2017). In part, we will expand here on some of the concepts discussed in rudimentary form at an Ernst-Strüngmann-Forum in 2015 in which the authors were involved (Kurth-Nelson et al., 2016). The fourth section will go into different (parametric and non-parametric) statistical methods for recovering neural and/or dynamical dynamics, or even identifying dynamical system itself, from neural and/or behavioral measurements. This article will then conclude with ideas on how to 'treat' maladaptive dynamics in the final section. For illustration of all DST concepts and ideas, we will use a simple recurrent neural network model which will provide a red thread throughout all sections. Although we aimed for a relatively non-formal presentation in general, and major take-homes should be accessible without any special mathematical background, we felt that a somewhat deeper understanding requires at least some formal concepts, which we will try to introduce gently.

## 2. Basics of DST

Formally, a dynamical system is defined by equations of motion which specify for a set of system variables how these change in time (and potentially along other dimensions like space). In continuous time, this gives rise to a system of differential equations

$$
\begin{aligned}
\dot{x}_1 &\equiv dx_1/dt = f_1(x_1, \ldots, x_M, t; \boldsymbol{\theta}) \\
\dot{x}_2 &\equiv dx_2/dt = f_2(x_1, \ldots, x_M, t; \boldsymbol{\theta}) \\
&\qquad\vdots \\
\dot{x}_M &\equiv dx_M/dt = f_M(x_1, \ldots, x_M, t; \boldsymbol{\theta})
\end{aligned}
\tag{1}
$$

where the $x_i$ are any dynamical variables which collectively describe the state of the system, for instance membrane voltages, neural firing rates, cognitive quantities, mood states or behavioral markers. $\boldsymbol{\theta}$ denotes parameters, for instance spiking thresholds, the strength of synaptic connections, or the strength of interactions between cognitive and emotional subsystems. Parameters in this context means these are taken to be fixed from the perspective of the dynamical variables $x_i$, but potentially they may themselves evolve according to dynamical rules on perhaps slower time scales, like, e.g., those governing synaptic long term plasticity. Dynamical system (1) is usually written more compactly in vector notation as

$$
\dot{\mathbf{x}} \equiv \frac{d\mathbf{x}(t)}{dt} = F_{\boldsymbol{\theta}}(\mathbf{x}, t).
\tag{2}
$$

In discrete time, a DS is given through a recursive map of the general form

$$
\mathbf{x}_t = F_{\boldsymbol{\theta}}(\mathbf{x}_{t-1}, \mathbf{s}_t),
\tag{3}
$$

where the $\mathbf{s}_t$ may represent time-varying external inputs. Since neuronal processes evolve in continuous time (while neural measurements, on the other hand, are always taken at discrete times), it may be important to point out that, in principle, discrete time can be transformed into continuous time formulations, and vice versa, by noting, for instance, that a continuous time DS can be obtained from a discrete time DS by taking the limit $\Delta t \rightarrow 0$

(4) $\quad \mathbf{x}_t = \mathbf{x}_{t-1} + \Delta t \cdot F(\mathbf{x}_{t-1}) \Rightarrow \frac{\mathbf{x}_t - \mathbf{x}_{t-1}}{\Delta t} = F(\mathbf{x}_{t-1}) \xrightarrow{\Delta t \rightarrow 0} \dot{\mathbf{x}} = F(\mathbf{x}).$

This is in fact the way continuous time systems are usually solved numerically on the computer, by translating them into an equivalent discrete time description that can be solved for time step by time step as on the left-hand side of eq. 4 (which is called the forward-Euler rule in this context; however, in practice, much more powerful, stable and efficient numerical solvers are used [e.g., Press, Teukolsky, Vetterling, & Flannery, 2007]). A more detailed discussion on how to derive a mathematically equivalent discrete time DS from a continuous time DS is given in (Ozaki, 2012) (in essence, if one is really aiming for mathematical equivalence, one has to take care that the change in $\mathbf{x}_t$ with each iteration of the recursive map (3) across some time step $\Delta t$ is exactly equal to the change in $\mathbf{x}(t)$ of the corresponding ODE system (2) obtained across the same time period $\Delta t$).

To continue, let us formulate a specific example along which to develop ideas and the discussion in the following sections. For this, we chose a simple but generic recurrent neural network (RNN) formulation commonly employed in both computational neuroscience and machine learning (e.g., Beer, 2006; Durstewitz, 2017b; Song, Yang, & Wang, 2016),

(5) $\quad \dot{\mathbf{x}} = -\tilde{\mathbf{A}}\mathbf{x} + \tilde{\mathbf{W}}\varphi(\mathbf{x}) + \tilde{\mathbf{L}}\mathbf{s}(t),$

or, similarly, in discrete time,

(6) $\quad \mathbf{x}_t = \mathbf{A}\mathbf{x}_{t-1} + \mathbf{W}\phi(\mathbf{x}_{t-1}) + \mathbf{L}\mathbf{s}_t.$

Note that here and in the following we do not aim for exact mathematical correspondence between these formulations – here these systems serve merely for illustration. $\phi(\cdot)$ is some nonlinear and usually, but not necessarily, monotonically increasing transfer or input/output function. Common choices are, for instance a sigmoid function $\phi(u) = [1 + \exp(-u)]^{-1}$, a piecewise linear function $\phi(u) = \max(u - h, 0)$ (or "rectified linear unit", ReLU) with 'firing threshold' $h$, or a non-monotonic radial basis function like the Gaussian. A biophysical interpretation of model (5)-(6) is given by considering variables $\mathbf{x}_t$ as membrane potentials (or currents) with passive decay time constants represented in diagonal matrices $\tilde{\mathbf{A}}^{-1}$ and $\mathbf{A}$, respectively, $\mathbf{W}$ as a matrix of synaptic connection weights, and $\phi(\cdot)$ as a membrane voltage to spike rate conversion function.

However, and importantly, although we have introduced this system as a RNN here, it can be thought of as a much more general description of a dynamical system which contains nonlinear 'nodes' $x_{it}$ interacting through coupling matrix $\mathbf{W}$, and with auto-regressive terms in $\mathbf{A}$. These nodes may as well represent behavioral variables, psychiatric symptoms, or individuals in a society. In fact, it could be shown that DS (6) is *universal* in the sense that it can approximate *any* truly underlying DS arbitrarily closely (under some mild conditions; Funahashi & Nakamura, 1993; Kimura & Nakano, 1998; Trischler & D'Eleuterio, 2016).

# 3. Dynamical phenomena and their potential relation to psychiatric conditions

## 3.1. Attractor dynamics and multi-stability

A comprehensive geometrical representation of a DS is its state space, which is the space spanned by all the dynamical variables of the system, as illustrated for a 2-unit RNN in Fig. 2B-D (for higher-than-3-dimensional DS, to obtain a state space-like visualization, we would need to perform some smart dimensionality reduction, or to look at various informative 2D or 3D projections of the system). A nice and powerful property of the state space representation is that, if all dynamical variables are given or known (as is the case for a computational model), it provides a complete description of the system's state, behavior, and (in the deterministic case) future fate: A point in this space completely (exhaustively) specifies the system's current state, and the so-called flow field (the arrows in Fig. 2B-D) completely specifies how the system will evolve in time when released at any point in this space (the flow field is obtained by simply plotting the temporal derivatives of eq. 5, or temporal difference vectors of eq. 6, within this space). The temporal evolution of the system's state within this space when started from a specific *initial condition* $\mathbf{x}_0$ (or $\mathbf{x}(0)$) is represented by its *trajectory* (black dashed line in Fig. 2B; Fig. 3A-C).

The state space in Fig. 2B highlights an interesting geometrical property of this 2-unit RNN: When released anywhere left from the green dashed line, following the flow field, the system's state always converges to a unique point in the upper left quarter of this space for $t \rightarrow \infty$. When released to the right from it, the system's state will always, inevitably, move toward another unique point in the lower right quarter. These two special points are called *fixed points* of the system dynamics, and, as indicated, lie exactly on the intersection of two specific curves in this space, the so-called *nullclines*. More formally, fixed points are points at which the system's state evolution would come to an absolute halt when exactly placed there, i.e. for which we have $\mathbf{x}_* = F(\mathbf{x}_*)$ (such that $\mathbf{x}_{t+1} = \mathbf{x}_t$) in the discrete, or $\dot{\mathbf{x}} = 0$ in the continuous time case (the flow will be exactly zero, no movement anymore). The two curves labeled as nullclines are the sets of points at which either one of the two system variables comes to a rest, i.e. for which we have $x_{1*} = f_1(x_{1*}, x_{2t})$ ($dx_1/dt = 0$) or $x_{2*} = f_2(x_{1t}, x_{2*})$ ($dx_2/dt = 0$) (more generally, therefore, there are as many nullclines as there are dynamical variables). Hence, the intersection of all these curves must give a fixed point. Fixed points can be *stable* or *unstable*, depending on whether activity *converges* toward these points (as for the two points just discussed), or whether it *diverges* from them along at least one direction, as is the case for the third fixed point in the center of the graph which also lies on an intersection of the two nullclines, and through which the green-dashed line in Fig. 2B runs. Stable fixed points are also called fixed point *attractors*, and they come with their *basins of attraction*, defined as the set of all points from which activity converges toward them – in fact, the dashed-green line divides the whole state space into the two basins of attraction corresponding to the two fixed point attractors.

If noise is added to the dynamical system (eq. 5, 6), it may cause trajectories to eventually cross the 'energy ridge' between attractors (Fig. 2E). The likelihood of such transitions or, conversely, the dwell times within specific states, will depend on the noise amplitude and the steepness of the attractor basins (Fig. 2C,D). This gives rise to a phenomenon called 'meta-stability' (Balaguer-Ballester, Moreno-Bote, Deco, & Durstewitz, 2017), where the underlying deterministic attractors may be truly stable, but noise-induced perturbations could terminate stable states and cause the system to hop around different attractor states (e.g., Fig. 2C).

Fixed points and attractors are neither mathematical curiosities nor metaphors, but real and common properties of physical and biological systems that can be described through sets of differential equations. The resting or steady state potential of a neuron, for instance, is a stable fixed point: After applying a small perturbation to the cell, for instance by current injection, the cell's membrane potential will always return to this stable base level. Likewise, mood may return to some stable set point after a small perturbation through an external event. Thus, again, we emphasize that the DST framework is equally applicable to phenomena on very different levels, although in the first example, a cell's resting potential, the phenomenon is easily accessible experimentally and can be neatly connected to the underlying biophysics, while in the second example, mood, the physical substrate is much less clear and proper experimental assessment more difficult.

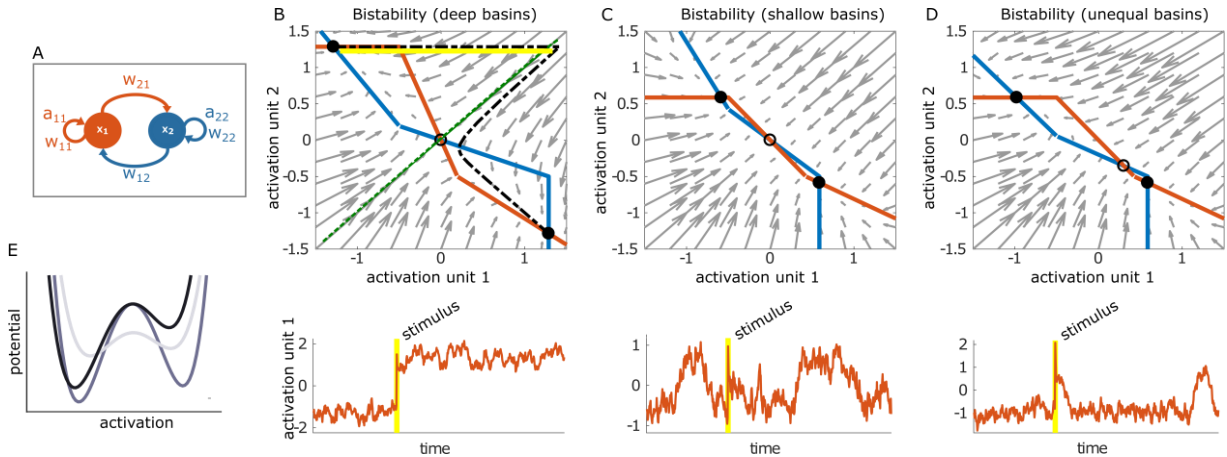

**Fig. 2. Example of multi-stability in a 2-unit PLRNN.** A) Two-unit toy PLRNN. B) depicts the flow field for the PLRNN in A (with parameters $a_{11} = a_{22} = .5$, $w_{11} = w_{22} = .36$, $w_{12} = w_{21} = -.36$, $h_1 = h_2 = -.5$), with gray arrows marking direction and magnitude of the flow. Blue and red lines are the nullclines of unit 1 (red) and unit 2 (blue), solid/open circles show stable/unstable fixed points, and the green line separates the basins of attraction of the two stable fixed points. The dashed black line shows the deterministic (noise-free) trajectory of the system, starting at the left FP, after a brief stimulus (yellow) to unit 1. C) Same as in B) with network parameters slightly changed, causing the system's attractors to move closer together and their basins to become shallower. D) Same as in B) and C), with parameters slightly changed such that the symmetry between attractor states is broken. The system now has one attractor with steeper basin than the other. The bottom parts of figures B)-D) show the activation in time of unit 1 for the systems above, a brief positive stimulus to unit 1 (yellow), and a small amount of noise ($\sigma^2 = .02$) added to the system. While in example B) the network is capable of upholding unit 1's high activation, remaining in the high-rate attractor despite the noise, C) shows how activity spontaneously switches between the two attractors due to the presence of noise. In D) the system only remains briefly in the high-rate attractor due to its small basin of attraction, from which it is kicked out by the noise after relatively short dwelling times. E) Schematic potential landscape depicting the extent and depth of the basins of attraction of the systems in B) (dark grey), C) (light grey), and D) (black). Potential minima correspond to the attractor states.

### 3.1.1. Multi-stability and line attractors in working memory, decision-making and inference

Dynamical systems may harbor not just one or two, but many different stable fixed points (or other attractor objects, see below). Such multi-stability, that is the co-existence of many attractor states, has been proposed to underlie functions like **working memory** (e.g., Durstewitz, Seamans, & Sejnowski, 2000b; Wang, 2001), with each fixed point corresponding to the active maintenance of a different memory item. The idea is that different briefly presented stimuli would push the network into one of the different stimulus-specific attractor states, which – by virtue of their attractor property – would maintain an active representation of the stimulus even after its removal that could guide action in forthcoming choice situations. Physiologically, the attractor states would correspond to elevated firing rates (termed 'persistent activity') in the respective subset of stimulus-memory-selective neurons, as supported by electrophysiological observations in, for instance, the dorsolateral prefrontal cortex (e.g., Funahashi, Bruce, & Goldman-Rakic, 1989; Fuster, 2015). Attracting dynamics are thought to play a role also in a variety of other cognitive processes (Albantakis & Deco, 2009; Machens, Romo, & Brody, 2005; Rabinovich, Huerta, & Laurent, 2008; Wang, 2008). Indeed, persistent activity and other indications of neural multi-stability are not at all unique to the dorsolateral prefrontal cortex, but have been observed in many other cortical, and even subcortical, structures such as orbitofrontal and anterior cingulate cortex, amygdala, primary visual cortex, and many others (Aksay, Gamkrelidze, Seung, Baker, & Tank, 2001; Miller, Li, & Desimone, 1993; Supèr, Spekreijse, & Lamme, 2001). In the hippocampus, attractor dynamics have long been related to **memory recollection and pattern completion** (Hopfield, 1982; Wills, Lever, Cacucci, Burgess, & O'keefe, 2005). Here, the idea is that a stimulus resembling one of the stable "memory" states will put the dynamical system within the attractor basin of that state. The dynamics will then lead the system to the deepest point of the basin of attraction, and thereby allow it to recover the 'typical' instance of that stimulus. A similar intuition has been used to apply dynamical systems to **decision-making**, with different attractors representing different choice options (e.g., Albantakis & Deco, 2009; Wang, 2002; Wang, 2008). For instance, the popular drift-diffusion decision-making account can be viewed as a stochastic dynamical process where the attractor strength determined by the stimulus drive and prior beliefs gives rise to the drift term, while the noise corresponds to the diffusion term (Bogacz, Brown, Moehlis, Holmes, & Cohen, 2006; Ratcliff, 1978; Ratcliff & McKoon, 2008). Noise, here, would generate the probabilistic component of decision-making.

Finally, at the extreme end, when attractor valleys become perfectly flat along one or more dimensions, we have a so-called **line-attractor** where the fixed points form a line, ring or plane (Durstewitz, 2003; Machens et al., 2005; Seung, 1996; Seung, Lee, Reis, & Tank, 2000), a continuum of *neutrally* stable fixed points along which there is neither con- nor divergence (see Fig. 4A). Thus, in the absence of noise, the system would just stick where put along this line, while noise would cause the system to slowly drift (friction-free) along this line (giving rise to a 'random walk'). Line attractors have been proposed to underlie phenomena such as parametric working memory (Machens et al., 2005) where a continuously valued quantity (like a 'flutter frequency' [Romo, Brody, Hernández, & Lemus, 1999] or spatial position [Zhang, 1996]) has to be retained in working memory. In parietal cortex, line attractor dynamics have been suggested to be involved in **inference and multidimensional integration** (Deneve, Latham, & Pouget, 2001). The intuition here is similar to that of parametric working memory in that the position along the line can be used to represent a continuously varying quantity. By carefully arranging the dynamics, this can be used for Bayesian optimal inference.

Multi-stability therefore constitutes a rather general dynamical phenomenon being put to use in qualitatively distinct functions based on the brain region within which it is expressed.

### 3.1.2. Modulators of attractor dynamics

Altering attractor dynamics may hence modify a variety of cognitive functions, from working memory to inference. Multiple ways of changing or modulating attractor dynamics can be distinguished. Flattening attractor basins will reduce the stability of goal or memory states in the face of new stimuli or noise fluctuations (Fig. 2C). Conversely, increasing the depth of attractor states will increase stability, but also reduce the ability to switch between representations, actions, or goals (Fig. 2B). Increasing the noise in the system will increase the likelihood that the system leaves an attractor and hops to a different state (Fig. 2C,E). Which attractor state the (neural) system assumes at any particular moment in time, and how quickly or easily it roves around between attractor states, hence depends on a number of factors, including the size, depth and separation of the basins of attraction, and the amplitude (variance) of noise.

There are numerous biophysical factors that could influence or modulate attractor basins and the amount of noise. At the biophysical level, probabilistic synaptic release (Stevens, 2003) or random channel state transitions (Koch & Laurent, 1999) can inject noise. Dopamine, via its synaptic and ionic actions, can regulate the width and steepness of basins of attraction (Durstewitz & Seamans, 2008), both in the prefrontal cortex (Durstewitz & Seamans, 2008; Durstewitz, Seamans, & Sejnowski, 2000a; Lapish, Balaguer-Ballester, Seamans, Phillips, & Durstewitz, 2015) and in the striatum (Gruber, Solla, Surmeier, & Houk, 2003). Stimulating dopamine D1 receptors is thought to render attractors more stable by – among other ionic effects – boosting NMDA-mediated recurrent excitation synergistically with $GABA_A$-mediated inhibition of competing assemblies (Durstewitz et al., 2000a), while D2 stimulation has the opposite effect (Durstewitz & Seamans, 2008). Modulators like dopamine could therefore affect the tradeoff between cognitive flexibility, supported by flat attractor basins that ease moving among representations, vs. working memory and goal orientation, facilitated by deep and wide basins that protect the current focus (Durstewitz & Seamans, 2008; Ueltzhöffer, Armbruster-Genç, & Fiebach, 2015). Similar roles have been suggested for serotonin (Maia & Cano-Colino, 2015) and NMDA-mediated disinhibition (Murray et al., 2014).

At the extreme, when neuromodulatory systems get out of balance, the dynamics may be influenced adversely, leaving a functional operating regime and leading to overly deep and sticky attractors, or – at the other extreme – impairing the stability of the system through overly flat attractors. Indeed, given that neuromodulators regulate cortical dynamics, and their importance in the aetiology and treatment of mental illnesses, it is tempting to speculate that a relationship might exist. We now turn to some examples of this.

### 3.1.3. Altered attractor dynamics in schizophrenia and ADHD

Schizophrenia is characterized by complex changes in dopaminergic function in both striatum and (frontal) cortex. Indeed, there has been extensive work on these systems which spans the biophysical, computational, physiological and neuro-behavioral levels (Durstewitz, Kelc, & Güntürkün, 1999; Durstewitz & Seamans, 2008; Durstewitz et al., 2000b; Floresco, Zhang, & Enomoto, 2009; Winterer & Weinberger, 2004; Yang, Seamans, & Gorelova, 1999). Building on data-driven biophysical models in conjunction with a dynamical systems account, Durstewitz and Seamans (2008) have argued that the increase in the affinity of the D2 receptor for dopamine seen across psychosis-inducing agents could lead to a flattening of attractor landscapes, while at the same time the reduction in prefrontal dopamine levels may cause overly deep attractor landscapes via preferential stimulation of D1-receptors during active task challenge. Both overly stable and very weak attractor states could cause working memory deficits, for different reasons and with different behavioral phenomenology, namely increased perseveration in the former vs. enhanced distractibility in the latter case (cf. Durstewitz & Seamans, 2008). At the same time, a flattening of attractor landscapes could also explain the experimentally observed deficits in cognitive flexibility (Armbruster, Ueltzhöffer, Basten, & Fiebach, 2012; Floresco, Block, & Maric, 2008). More recent work examining the effect of distractors on working memory furthermore suggests that also ketamine, and other potentially schizophrenia-related biophysical changes, may contribute to reduced stability of attractor states (Murray et al., 2014; Starc et al., 2017). Specifically, weakening the lateral inhibitory connectivity (via reduced NMDA drive on interneurons) results in broader attractor states (see also Durstewitz et al., 2000b; Murray et al., 2014); this correctly predicts sensitivity to more distant distractors, which in turn is related to the temporal drift in working memory.

Weak and flat attractor landscapes, and the distractor susceptibility that comes with it, may also lead to incoherent and disorganized thought as the network state meanders around mainly driven by the system noise, or may lead to spontaneous pop-out of representations experienced as hallucinations (Durstewitz & Seamans, 2008). Thus, the depth of attractor landscapes might relate to the ability to integrate information into a stable mental representation of an over-arching topic or goal that organizes and guides specific cognitive processes in its service. This may also provide a dynamical systems account for the kinds of deficits seen in attention deficit hyperactivity disorder (ADHD; Forster & Lavie, 2016; Wilens, Faraone, & Biederman, 2004). However, high distractability and the inability to shield internal representations from external perturbations may not only be the result of weak attractors, but could also – or in addition – have their origin in other dynamical effects, like very high intrinsic noise levels or too strong (and possibly unselective) inputs into the system (Bubl et al., 2015; Cortese et al., 2012; Hauser, Fiore, Moutoussis, & Dolan, 2016). Different such dynamical systems-level explanations may have quite different implications for treatment. It may therefore be important to find out which of these dynamical possibilities underlies the deficits observed at the cognitive level. This is, in principle, feasible: attractor depths, noise levels, and input strength are all, in theory, distinguishable from experimentally observed neurophysiological time series (see sect. 4).

### 3.1.4. Strong attractors in ruminative preoccupations

Incessantly recurring obsessions or compulsions as typical for Obsessive Compulsive Disorder (OCD), on the other hand, and the inability to distance oneself despite cognitive effort (Markarian et al., 2010), may be explained in terms of overly strong attractors for such thought or action sequences (see e.g., Rolls, Loh, & Deco, 2008). At the physiological level, this increased attraction could be related to higher glutamate levels (Bhattacharyya et al., 2009; Pittenger, Bloch, & Williams, 2011) which may amplify recurrent excitation and thus attractor strengths. A related phenomenon might occur in dissociative or catatonic states, where external stimuli are seemingly unable to change the state of the person. This would be characteristic of a system that is locked in a deep attractor state. In dissociative phenomena in posttraumatic stress disorder, these deep prefrontal attractor states might concomitantly inhibit bottom-up sensory drive (Lanius, Brand, Vermetten, Frewen, & Spiegel, 2012; Lanius et al., 2010; Sack, Cillien, & Hopper, 2012).

Beyond working memory, fixed-point-like attractor dynamics, as it may emerge from Hebbian plasticity in neural networks, has served more generally as a model of memory formation and recollection (Hopfield, 1984). Memories for emotional stimuli are altered in a variety of disorders. In anxiety disorders, for instance, overgeneralization of fear is a phenomenon whereby the fear response carries over from threatening stimuli to relatively distantly related cues (Lissek & van Meurs, 2015; Thome et al., 2017). This misattribution of fear has been understood as an enhanced integration of stimuli into the fear network (Ehlers & Clark, 2000; Lang, 1988). One way to explain this could be a widening of the memory attractors arising from fear learning.

Strong attractor states in emotion and self-referential processing systems (Nejad, Fossati, & Lemogne, 2013), as they may occur for instance during rumination in major depressive disorder (MDD), may at the same time destabilize attractor states in systems which mediate cognitive processes (Ramirez-Mahaluf, Roxin, Mayberg, & Compte, 2015). This idea is illustrated in Fig. 2B-D with two units (which one may think of as representing two network hubs in this context) with strong self-excitation but mutual inhibition. As illustrated, by either increasing the amount of self-excitation in one of the two hubs or through an imbalance in the feedback between the two (Fig. 2D), one of the two attractor basins may strongly expand at the expense of the other. Such a configuration could account for symptoms such as rumination and negative mood that occur concurrently with lack of attention and impaired decision-making (Gotlib & Joormann, 2010; Lyubomirsky, Kasri, & Zehm, 2003).

An interesting aspect to point out is that certain feedback loops in the brain may be much more sensitive to biophysical parameter changes or therapeutic 'inputs' than others (e.g., dopamine affects a whole variety of different intrinsic and synaptic channels across many brain regions). In devising effective pharmacological or behavioral treatments it would thus be important to not only determine how they would alter attractor landscapes, but also which would be the most effective knobs to turn.

## 3.2. Sequential phenomena: limit cycles and saddle node channels

Fig. 3A illustrates another setup of the RNN defined through eq. 6. A slight change in some of the system parameters (cf. Fig. 3 legend) gives rise to a different set of phenomena: Rather than converging to a stable fixed point, the RNN now continues to periodically oscillate. It is not a simple harmonic (sinusoidal-type) oscillation, however, but a more complex waveform that is repetitively produced. This complex but still periodic waveform represents another type of attractor state, termed a stable *limit cycle* (just as with fixed points, there are also unstable limit cycles). Limit cycles can become quite complicated in appearance, with multiple different minima and maxima and very long periods in which the system's trajectory does not precisely retrace itself, although they always close up eventually. Limit cycles often result from interacting positive and negative feedback loops, as ubiquitous in the nervous system, and may represent sequences that are to be repetitively produced, like potentially complex, but still relatively stereotypical motor programs or movement patterns. For instance, the activity of central pattern generators in motor nuclei or ganglia that generate repetitive running or swimming movements (Kato et al., 2015; Marder & Bucher, 2001; Marder, Goeritz, & Otopalik, 2015) or patterns in primate motor cortex (Russo et al., 2018) may correspond to limit cycles. Stereotypical, repeating movement patterns are observed in some psychiatric conditions (Ridley, 1994; Turner, 1999), and – at a higher cognitive level – perseveratively reoccurring chains of the same thoughts may potentially be described this way. In general, (nonlinear) oscillations – the equivalent of limit cycles in the time domain – are a hallmark of nervous system activity (Buzsáki & Draguhn, 2004), and specific alterations for instance in the gamma or delta frequency band have indeed been described in schizophrenia (Uhlhaas, Haenschel, Nikolic, & Singer, 2008) or ADHD (Demanuele et al. 2009). Just as with fixed points, there could also be multi-stability among limit cycles, or among limit cycles and fixed points (Fig. 3A).

Another psychiatrically relevant domain where limit cycles may arise is from the effects of mood states on reward learning (Eldar & Niv, 2015; Eldar, Rutledge, Dolan, & Niv, 2016). If moods modulate reward and loss of sensitivity and thereby drive expectations (Huys, Pizzagalli, Bogdan, & Dayan, 2013; Nusslock, Young, & Damme, 2014), but expectations in turn drive moods on a slower timescale, then this feedback loop can set up an oscillatory cycle where mood will periodically fluctuate. Eldar and colleagues showed that this may be more pronounced in individuals with more unstable mood patterns. Transitions among fixed point attractors caused by a limit cycle which itself develops on a much slower time scale are a common phenomenon in the nervous system (e.g., Durstewitz & Gabriel, 2007; Rinzel & Ermentrout, 1998): The slowly varying system variables mediating the limit cycle (mood states in this case) may drive a faster and by itself multi-stable sub-system (related to stimulus evaluation here) repeatedly across bifurcation points (see sect. 3.5, cf. Fig. 4D) as the oscillations wax and wane. Thereby this sub-system, which lives on a much faster time scale, will be driven back and forth between various of its 'attractor states' (to be precise: these would not be true attractor states of the *full* system including the slow variables, but they would be from the perspective of the faster sub-system if the slow variables were considered fixed parameters, an approach called 'separation of time scales' [Rinzel & Ermentrout, 1998; Strogatz, 2018]). More generally, stable limit cycles may ultimately underlie many types of symptom clusters which emerge in periodic intervals.

Of note, whether a dynamical phenomenon *observationally* manifests as a fixed point or as a limit cycle also depends on the variables measured and the level of description: Ongoing regular spiking behavior of a neuron, for instance, may constitute a stable limit cycle at the level of the *membrane potential*, but may appear as a stable fixed point in terms of the neuron's *firing rate*. Formally, this relation can be expressed, for instance, in terms of so-called Poincaré maps (specifically chosen sections through a system's state space), where (un)stable limit cycles of the full system would correspond to (un)stable fixed points of the map.

There are also other, more flexible ways to generate sequences in dynamical systems, as illustrated in Fig. 3B (Rabinovich, Huerta, Varona, & Afraimovich, 2008; Rabinovich, Varona, Selverston, & Abarbanel, 2006). Here, orbits connect a chain of saddle points, that is, 'half-stable' fixed points towards which activity converges from many directions but leaves along others. The curves that connect the different saddle points are called 'heteroclinic orbits' (Fig. 3 B), and the whole arrangement of heteroclinic orbits connecting a chain of saddle points has been termed a 'heteroclinic channel' (HC; Rabinovich, Huerta, Varona, et al., 2008; Rabinovich et al., 2006). The HC acts like an attractor, pulling in trajectories from the vicinity which then travel along the curves connecting the saddle points. In the completely noise-free system, activity would get stuck in the saddle points, but a little bit of noise is sufficient to push the system's state again onto the saddle point's unstable manifold, that is the heteroclinic orbit which connects it to the next fixed (saddle) point. Thus, like within a stable limit cycle, the system moves through a continuous (or discrete) series of states that may implement a sequence of thoughts or actions. There are at least two central features, however, where this constellation pronouncedly differs from limit cycles: First, the HC is not necessarily automatically repeating – it may start and terminate in a stable fixed point (as in the example in Fig. 3B). Hence, the system would move through the whole sequence only *once* when triggered by an internal or external stimulus. Second, this arrangement is much more flexible: Saddle points may be removed from or added to the already existing sequence through proper changes in the system parameters. Because of these properties, Rabinovich, Huerta, Varona, et al. (2008) advertised this framework as a much more plausible dynamical systems perspective on higher cognitive functions (e.g., syntactical sequences) than would be provided through limit cycles or fixed point attractors alone.

Whether psychiatric phenomena like the progression through different stages in schizophrenia or bipolar disorder are due to limit cycles, heteroclinic channels, or – for instance – noise pushing the system around among several meta-stable states, is a hard but empirically addressable question. Different phenomena come with different spatio-temporal signatures (see sect. 4); for instance, true limit cycles would usually be associated with one or a few relatively sharp peaks in the power spectrum, unlike meta-stable transitions. In general, determining which of a range of different dynamical systems phenomena best accounts for a set of behavioral or neural observations is, however, non-trivial, and requires detailed quantitative examination (cf. sect. 4), rather than inferences drawn from just superficial similarities (e.g., inferring a limit cycle from roughly cyclic behavior). This is particularly important as the specific type of dynamical mechanism could have important implications for optimal treatment, e.g. for both the type and timing of pharmacological medication.

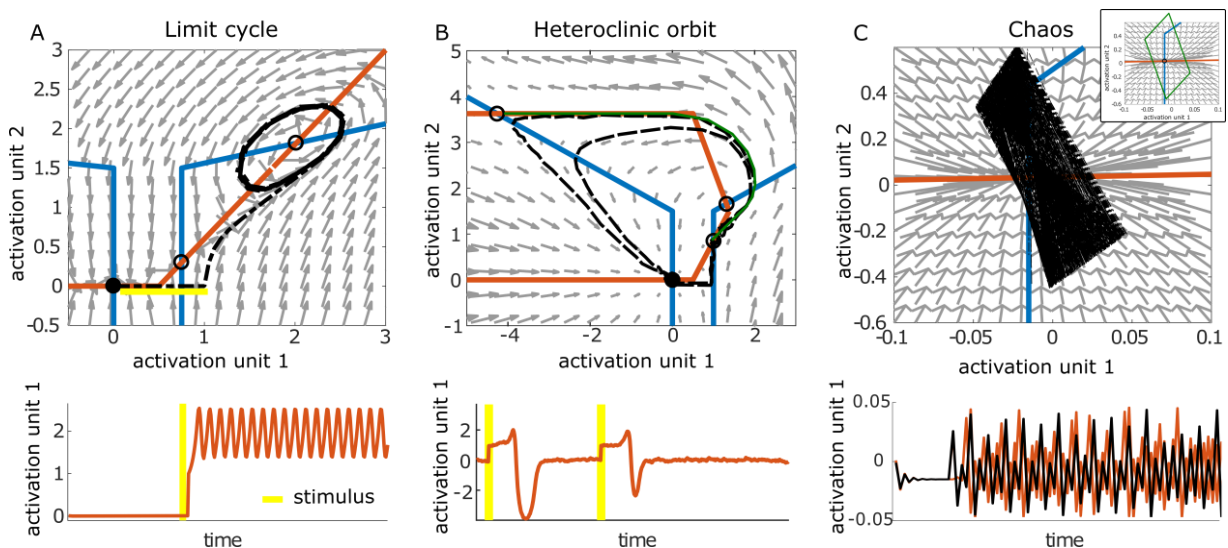

**Fig. 3. Examples of different sequential phenomena in dynamical systems, illustrated with the 2-unit PLRNN.** Panels depict flow fields (top row) and time graphs of unit 1 (bottom row) for slightly different parameter settings of the 2-unit PLRNN. Red and blue lines in flow fields mark nullclines of units 1 and 2, black dotted lines show one trajectory starting from (0,0), yellow lines mark brief presentations of positive external stimuli. A) Bistability among a stable limit cycle ($a_{11}$=.9, $a_{22}$=.5, $w_{11}$=.3, $w_{22}$=0, $w_{12}$=-.8, $w_{21}$=.6, $h_1$=.5, $h_2$=1.5) surrounding the right unstable fixed point, and a stable fixed point (on the left). A brief stimulus to unit 1 takes the system from its stable fixed point to the stable limit cycle. B) Heteroclinic orbit (shown in green) connecting the system's two saddle nodes ($a_{11}$=.8, $a_{22}$≈.77, $w_{11}$= $w_{22}$≈.4, $w_{12}$≈-.4, $w_{21}$=.4, $h_1$=.5, $h_2$=1.5). In this case, the heteroclinic channel (HC) created by this orbit is itself not attracting (in contrast to examples in Rabinovich et al. (2008)), such that nearby trajectories tend to diverge from it (but note that, in the deterministic case, the system would continue to move on the heteroclinic orbit if placed exactly on it). Yet, this unstable HC still influences the behavior of the system in the sense that brief perturbations through an external stimulus (shown in yellow) tend to cause trajectories to move slowly in its vicinity (just below it) until they return to the stable fixed point at the bottom. C) Chaotic attractor ($a_{11}$≈.46, $a_{22}$≈-1.2, $w_{11}$≈.09, $w_{22}$≈2.47, $w_{12}$≈.32, $w_{21}$≈.28, $h_1$≈-.27, $h_2$≈.43). The system state is drawn initially into the vicinity of an unstable fixed point, but is then repelled away from it and starts moving chaotically within a bounded region surrounding this unstable fixed point (marked approximately by green quadrangle in the top inset graph). In fact, as characteristic of chaotic systems, two very close-by initial conditions may lead into very different activation patterns in the longer run, as displayed for unit 1 in the bottom graph. Note that for chaos to occur in an ODE system, eq. 1-2, at least three dimensions (i.e., dynamical variables) are required, while chaos in a nonlinear recursive map, eq. 3, only needs one variable. The PLRNN, as used as a basis for these simulations, is a 2-variate map. In A) and C), flow vectors are all normalized to same length for better visualization of flow direction.

## 3.3. Attractor ghosts and the regulation of flow

Another important phenomenon in dynamical systems is that of 'attractor ruins' (Tsuda, 2001, 2015), also termed 'attractor ghosts' (Strogatz, 2018), quasi- or semi-attracting states (Balaguer-Ballester, Lapish, Seamans, & Durstewitz, 2011), or, relatedly, 'slow points' (Sussillo & Barak, 2013; the terms 'attractor ruins/ghosts' are more general in that an attractor ruin does not necessarily have to come from a single point but could be due, e.g., to a limit cycle). These are 'attractors' which just lost their stability (in fact, just ceased to exist), i.e. *almost* stable objects (or regions) in the system's state space to which trajectories still converge along most directions, but from which some trajectories will slowly escape along others (Fig. 4B,C). Thus, although formally, strictly, there is no fixed point, limit cycle, or chaotic attractor (see below), in the system's *parameter* space we are at least *very close* to one, i.e. we are just *on the edge* of a bifurcation (see below) where with only a slight change in parameters one of these objects would come into existence (e.g., Fig. 4D).

This comes with important and interesting implications that differentiate these objects from either true attractors or clearly unstable objects. One of these properties is that trajectories are still driven toward these objects from a larger region of state space, in that sense still forming a kind of basin of attraction. Another is that trajectories considerably *slow down* in the vicinity of these objects, i.e. they tend to prevail in attractor ruins (formally this could be understood by noting that while within a fixed point the flow is exactly zero, close to points which are not quite fixed points the flow would still be close to zero if the system is defined by a *smooth* and *continuously* changing map or flow). Just as with heteroclinic channels, this phenomenon could be exploited for flexible sequence generation with trajectories traveling among attractor ruins while returning to some stable base state in the end (Russo & Treves, 2012). It should be noted that there are several related phenomena that may be empirically hard to distinguish: Rather than a fixed point that vanishes and thus becomes an attractor ghost, for instance, it might be at the edge of changing from stability to instability (another type of bifurcation). This means the flow is close to changing sign from converging to diverging in the fixed point's vicinity, and hence will effectively slow down as we approach that instant in a smooth system (although the fixed point itself never vanishes; e.g., Durstewitz, 2017b).

Another interesting aspect here is that configurations close to line attractors, slow points or attractor ruins can introduce new effective time constants into the system which could be largely independent from the system's intrinsic time constants (like those of the neuronal membrane or synaptic transmission). Moving farther from or closer to a true attractor state in parameter space, the time it takes for the system to travel through the attractor ruin is controlled (going to infinity as we get arbitrarily close) – this phenomenon has been exploited, for instance, for interval timing in neural systems (Durstewitz, 2003, 2004). It might also account for timing problems evident in Parkinson's disease (Rammsayer & Classen, 1997) or ADHD (Rubia, Halari, Christakou, & Taylor, 2009), given that the dopaminergic system has been linked to alterations in (interval) time perception and production (Hass & Durstewitz, 2014; Rammsayer, 1993). In light of the ideas discussed in the previous section on hyper-stable mood-related attractor states that de-stabilize attractors in other systems, slow thinking and slowed-down mental processes, as commonly observed cognitive symptoms in MDD patients, could potentially be caused by a similar phenomenon: If attractor basins become very flat and shallow, close to line and 'plane' attractors which have perfectly flat directions in state space (Fig. 4A), processes operating on this landscape, such as thought or movement programs, would evolve very slowly.

## 3.4. Chaos

With a further change in the PLRNN parameters, strange things start to happen. Although we are with the completely deterministic case, no noise added, the system's activity appears irregular and somewhat stochastic (Fig. 3C). This *aperiodic* and *irregular* yet completely deterministic behavior, with the system's states never quite repeating themselves, is called *chaos* (Ott, 2002; Strogatz, 2018). The state of chaos can still be an attractor, pulling in trajectories from a larger basin of attraction into a bounded region of state space within which trajectories would, if unperturbed, continue to travel forever. They would not form a closed orbit, however, as with a limit cycle, but would continue to move along new paths, yet, while doing so, would stay within a bounded set or volume (Strogatz, 2018). Unlike fixed point attractors which have only converging directions, and stable limit cycles toward which trajectories converge from all directions except on the limit cycle itself where it is neither con- nor diverging, chaotic attractors have at least one direction along which trajectories *diverge*, but in such a way that they are 'reinjected' into that same volume that encloses the chaotic attractor (Strogatz, 2018). Because of this divergence, activity on the chaotic attractor is highly (exponentially) sensitive to perturbations and minimal differences in initial conditions, with small changes resulting in quite different trajectories (Fig. 3C bottom). The fact that activity in chaotic attractors is (highly) irregular yet not random, thus retaining a certain degree of serial predictability that, for instance, a white noise process lacks, may be beneficial for certain cognitive and coding purposes (Durstewitz & Gabriel, 2007). In a sense, it creates deterministic variation around a central theme which may be relevant to cognitive search and creativity (Schuldberg, 1999; Zausner, 1996). Chaotic objects could also be unstable (meaning that trajectories fully diverge from them, in an irregular pattern, and do not return). Especially interesting from a computational perspective is the phenomenon of *chaotic itinerancy* (Tsuda, 2001, 2015) where trajectories chaotically traverse between different attractor ruins, a setup that has been exploited for dynamic and flexible sequence production and recognition (Russo & Treves, 2012). There are a number of further interesting dynamical phenomena related to chaotic activity, like *intermittency* (where long periods of silence or almost regular activity are interrupted by chaotic bursts), or highly complicated and fractal *riddled* basins of attraction (Ott, 2002). Another deterministic phenomenon, *quasi-periodicity* (Strogatz, Friedman, Mallinckrodt, & McKay, 1994; Strogatz, 2018), is sometimes difficult to distinguish from chaos because it is also associated with never-repeating trajectories, although these are much more regular than in a truly chaotic system.

It seems that placing a neural system at the edge of chaos, or slightly into a chaotic regime, is highly beneficial from a computational perspective (Bertschinger & Natschläger, 2004; Jaeger & Haas, 2004; Legenstein & Maass, 2007; Sussillo & Abbott, 2009): In this regime the system naturally expresses complex temporal structure while at the same time hanging on to external stimulus information for a while. In contrast, if the system is too regular (too convergent) it exhibits no interesting internal behavior, while if it is too chaotic (too divergent) it quickly forgets about external stimuli and behaves too 'randomly'. Therefore, it appears that a slightly chaotic ground state is optimal for performing a variety of computational operations (Bertschinger & Natschläger, 2004; Legenstein & Maass, 2007; Sussillo & Abbott, 2009). Consequently, if the brain leaves this computationally optimal regime and migrates either too much into the regular or too much into the chaotic range, cognitive and emotional problems may ensue. Indeed, PTSD patients, for instance, show a highly reduced heart rate variability (i.e., larger regularity), which is assumed to be an index of a reduced ability to flexibly respond to incoming information (Thayer, Ahs, Fredrikson, Sollers, & Wager, 2012; Thayer & Lane, 2000). Likewise, diminished variability in mental states has also been described in higher age (Battaglia et al., 2017). On the other hand, a highly chaotic regime with its sensitivity to perturbations may account for attentional problems and a high distractibility by, or over-sensitivity to, external stimuli, as, e.g., observed in ADHD. Thus, ADHD patients may suffer less from too large noise in the system, as discussed in sect. 3.1.3, but from a too chaotic internal dynamics. Again, for treatment, it may be important to find out which of these two possibilities provides the better account for the symptoms, although in practice noise and chaos are hard to disentangle (but see, e.g., Durstewitz & Gabriel, 2007).

## 3.5. Phase transitions and bifurcations

In the discussion of phenomena in dynamical systems above, we have repeatedly highlighted that many of these phenomena may be obtained within the very same system (Figs. 2-4), just by changing some of its parameters. This gives rise to another highly important observation: As system parameters are smoothly changed, we may often encounter dramatic and abrupt, *qualitative* changes in the system's behavior at some critical point (Fig. 4D). These are points in parameter space, also called *bifurcation* points (with *phase transitions* being special kinds of bifurcations), where the set of dynamical objects and/or their properties change, i.e. where certain fixed points, limit cycles or chaotic objects either come into existence, vanish, or change stability. For instance, as illustrated in Fig. 4D, as a system is smoothly driven across a bifurcation point, e.g. by slightly increasing synaptic weights, the system may suddenly hop into a remote region of state space because the attractor state it was just in has disappeared or became unstable. The kind of 'gradual' and 'linear' thinking we are often used to in our everyday lives, may thus completely break down in complex (or even quite simple) nonlinear dynamical systems such as the brain.

This observation in dynamical systems is likely to have profound implications for our understanding of crucial transitions, sudden onsets or offsets, or different distinct phases in psychiatric illnesses. That neural systems may undergo critical bifurcations with dramatic consequences is comparatively well established in epilepsy (Jirsa, Stacey, Quilichini, Ivanov, & Bernard, 2014; Naze, Bernard, & Jirsa, 2015), where one has relatively clear electrophysiological signatures that allow to identify and distinguish different types of bifurcations (see also Izhikevich, 2007). Indeed, at a more cellular and network-physiological level, bifurcations between different dynamical regimes (e.g., from regular spiking to bursting to chaos) have been quite well described experimentally where they can be provoked by changes, for instance, in the ambient level of NMDA receptor activation or other transmitter systems and intrinsic ion channels (Brunel, 2000; Durstewitz, 2009; Durstewitz & Gabriel, 2007; Izhikevich, 2003; Izhikevich, 2007; Naze et al., 2015; Rinzel & Ermentrout, 1998).

At a more cognitive level, bifurcations may account for sudden transitions observed in behavioral choices and the accompanying neural activity during the learning of a new rule (Durstewitz, Vittoz, Floresco, & Seamans, 2010). Also, both brief amnestic periods (Spiegel et al., 2017), during which stored memories cannot be recalled, as well as so-called lucid moments in dementia (Normann, Asplund, Karlsson, Sandman, & Norberg, 2006), where suddenly mnemonic details can be recovered again that appeared to have been lost, suggest that neural systems may sometimes hover at the edge of a bifurcation across which they might be driven back and forth by slight biophysical parameter changes. Bifurcations may also help to explain why psychopharmacological treatment sometimes helps and in other instances completely fails: Ramirez-Mahaluf et al. (2015), for instance, mimicked the effects of increased glutamate reuptake and selective serotonin reuptake inhibitors (SSRI) on network activity, and found that while within a certain range 'healthy' attractor dynamics could be pharmacologically restored, especially after passing critical bifurcation points network changes appeared irreversible by pharmacological means, potentially explaining non-responsiveness to drugs. In such cases, or more generally to kick a neural system out of particularly deep attractor states, more profound perturbations such as electro-convulsive therapy (ECT) or deep-brain stimulation (DBT) may be required (Mayberg et al., 2005; UK ECT Review Group, 2003).

Evidence for attractor transitions indicative of a bifurcation has recently also emerged from the application of DST to experience sampling, where emotions, events and at times cognitions or interpretations, are assessed multiple times per day over a period of a week or longer. Different emotions, it turns out, are not just simple responses to external factors, but influence each other directly, with sadness for instance increasing anger and anxiety but decreasing happiness (Bringmann et al., 2016). More interestingly, the change between states of depression and health shows signatures that are typical of a phase transition, namely so-called critical slowing down (similar as in Fig. 4B) where the autocorrelation of different emotions increases (van de Leemput et al., 2014).

In summary, bifurcations are a very common phenomenon in nonlinear systems such as the brain. A crucial and important aspect about bifurcations is that the system dynamics changes *qualitatively* and abruptly, sometimes quite dramatically, as a parameter of the system gradually changes. In other words, a potentially quite (infinitesimally) small change in a parameter may lead to a completely different behavior of the system, a somewhat non-intuitive feature of bifurcations. From this perspective, one may see therapeutic efforts in psychiatry as attempts to prevent certain bifurcations from happening or to induce others.

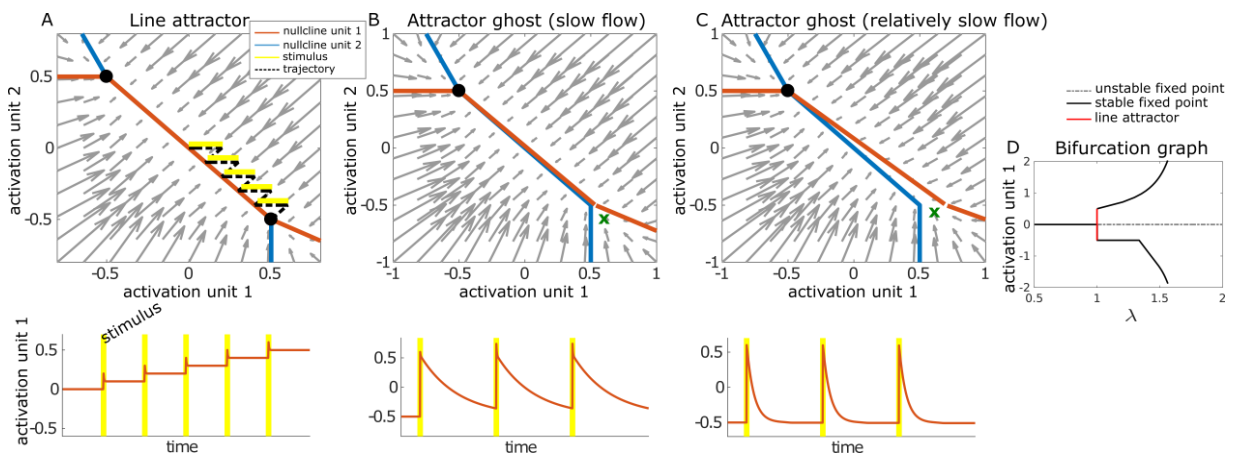

**Fig. 4. Example of line attractors, slow flow, and bifurcations in a 2-unit PLRNN.** A), B), and C) depict flow fields for slightly different parameter settings of the 2-unit PLRNN. A) shows a line attractor (with parameters $a_{11}=a_{22}=.4$, $w_{11}=w_{22}=.3$, $w_{12}=w_{21}=-.3$, $h_1=h_2=-.5$). Red and blue lines mark the nullclines of units 1 and 2. Black dotted lines show one trajectory starting from (0,0) and briefly pushed away from the line attractor by stimulus pulses (indicated in yellow). When the parameters of this line attractor configuration are changed (by reducing $w_{12}$ slightly), the system's

bottom-right fixed point disappears and leaves behind an 'attractor ghost'. In the vicinity of this attractor ghost the flow is very slow (B) or relatively slow (C), depending how far the system's parameters were moved away from the truly attracting configuration. The bottom figures show the activation of unit 1 for systems in A), B), and C), respectively, starting from the left-most attractor (for B and C), or at (0,0) (for A)), and stimulated repeatedly as indicated by the yellow lines. Note that for illustrative purposes we omitted trajectories and stimuli from B) and C). The stimuli in B) and C) take the state of the system to the spots in state space indicated by the green crosses. D) Example of a bifurcation for the system in A) with fixed points plotted as a function of network parameter $\lambda$. Here, $\lambda$ is a factor which regulates the size of the auto-regressive terms of all units in matrix $\mathbf{A}$ (see eq. 6). With $\lambda < 1$, the network exhibits only a single stable fixed point, then switches to a line attractor for $\lambda = 1$ (as in A), and finally harbors two stable and one unstable fixed points for $\lambda > 1$ (as in the systems in Fig. 2).

# 4.  Inferring DST objects from experimental data

As we had stressed in the Introduction, system dynamics and attractors are not merely theoretical concepts used to gain an abstract, phenomenological understanding, but they are 'real' in the sense that they could be extracted from experimental data. Broadly, one may think of 'bottom-up' approaches which investigate network-dynamical consequences of biophysical or biochemical manipulations in simulated neural systems, and 'top-down' approaches which aim to directly infer the system dynamics from observed time series data. Here we will focus on this latter approach, but will briefly review the former as well. Moreover, the approaches we will discuss here aim to reconstruct a system's state space and dynamics *directly* (and, ideally, completely). In contrast, properties of the underlying dynamical system may also be inferred less directly from power spectra, power or other scaling laws (Jensen 1998, but see Nonnenmacher, Behrens et al. 2017), perturbation approaches (Aksay et al., 2001), change points (Durstewitz et al., 2010), or a variety of other measures and properties of the observed time series (e.g., Niessing & Friedrich, 2010; Wills et al., 2005).

In the bottom-up approach, one sets up biologically usually quite realistic model networks which incorporate biophysical, biochemical, and structural (morphological, anatomical) details directly derived from experimental studies. For instance, at a biophysically very detailed level one may describe each single model neuron by a set of hundreds to thousands of differential equations which specify current flow throughout a realistically modeled somato-axonal-dendritic structure, and across the cellular membrane, generated by an array of passive, voltage- and chemically-gated membrane conductances in series with different ionic batteries (usually $Na^+$, $K^+$, $Ca^{2+}$, $Cl^-$; [Markram et al., 2015; Traub et al., 2005]). At a somewhat more abstract level, one may still set up single neuron models with parameters systemically inferred from physiological measurements (e.g., Hertäg, Hass, Golovko, & Durstewitz, 2012), and connect them into anatomically realistic columnar and layered architectures that are able to *quantitatively* reproduce *in vivo* electrophysiological recordings (Hass, Hertäg, & Durstewitz, 2016).

Due to their closeness to the real physiological substrate, such systems allow to translate measurements from single ion channels, neurons, or synapses, fairly directly into the model framework. For instance, based on a detailed characterization of the effects of dopamine D1R- and D2R-class stimulation on a variety of different voltage- or synaptically gated ion channels in prefrontal cortex (PFC) slices *in vitro* (Durstewitz & Seamans, 2002, 2008; Durstewitz et al., 2000a; Seamans, Durstewitz, Christie, Stevens, & Sejnowski, 2001; Seamans, Gorelova, Durstewitz, & Yang, 2001; Yang & Seamans, 1996), these authors used biophysical model networks to deduce the impact of these physiologically observed alterations on attractor dynamics in PFC (see also Brunel & Wang, 2001; Compte, Brunel, Goldman-Rakic, & Wang, 2000). From these model simulations, the lump effect of the D1R-induced changes appeared to be a widening and deepening of the basins of attraction associated with the different active memory states, while D2R-induced changes, on the contrary, caused a flattening of attractor basins, facilitating transitions among attractor states. This led to the 'dual-state theory' (Durstewitz & Seamans, 2008) which proposes a D1R-dominated regime beneficial for working memory and goal state maintenance, and a D2R-dominated regime facilitating cognitive flexibility, with the balance between the two regulated by dopamine concentration (Seamans, Gorelova, et al., 2001; Ueltzhöffer et al., 2015). A disturbance of this balance as in schizophrenic patients or ADHD (Meyer-Lindenberg & Weinberger, 2006) may consequently explain some of their cognitive symptoms.

What we called a 'bottom-up' approach above incorporates crucial physiological and anatomical details into a computational model which is then used to infer consequences for the dynamics. But there are also techniques for inferring system dynamics and attractor states directly from observed data, both fully model-free ('non-parametric') methods, as well as methods that build on a kind of symbiosis with generic dynamical systemmodels. One completely model-free approach to this problem is based on the idea of a temporal delay embedding of an experimental time series (Kantz & Schreiber, 2004; Sauer, Yorke, & Casdagli, 1991; Takens, 1981, Fig. 5). Say we have access to just one variable $x_t$ of a much higher-dimensional *deterministic* dynamical system in which all variables are coupled (as in a recurrent neural network), then the original attractor can be topologically faithfully reconstructed from this one variable by building a vector space out of time-lagged versions of this variable, $\mathbf{x}_t=(x_t, x_{t-\Delta t}, x_{t-2\Delta t}, \ldots, x_{t-(m-1)\Delta t})$ (Fig. 5). If the dimensionality $m$ of this 'delay-embedding space' is large enough (formally, larger than twice the attractor's so-called box-counting dimension, see Kantz & Schreiber, 2004), the delay-embedding theorems (Sauer et al., 1991; Takens, 1981) ensure us that trajectories within this reconstruction space will have a 1:1 correspondence to those in the original (only partially observed) space. Intuitively one may understand this result as follows: If we have observed only one variable $x_t$ of a $N$-dimensional (with $N>1$) dynamical system at time $t$, then there are (possibly infinitely) many ways one could have arrived at this one particular measurement – the situation is highly ambiguous. As we incorporate more and more of $x_t$'s history into the current representation, however, we are placing more and more constraints onto the underlying system's temporal evolution and thus start to disambiguate it until, at least in the deterministic case, we ultimately end up in a trajectory that can come only from one particular type of attractor dynamics.

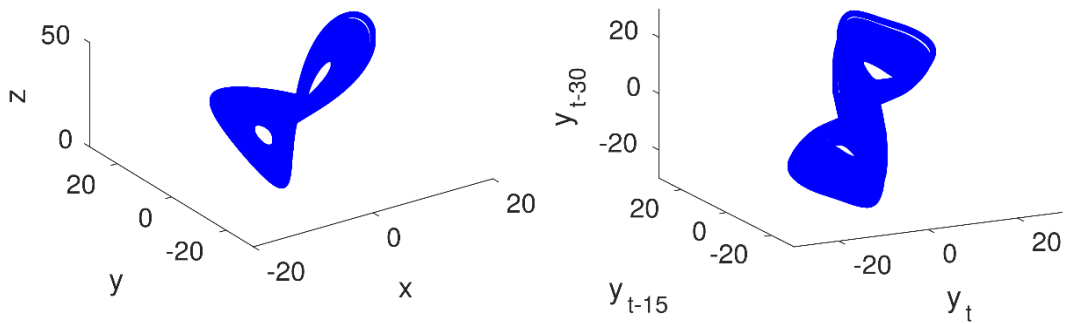

**Fig. 5. Illustration of the delay embedding technique.** Left: true trajectory of the system (chaotic Lorenz attractor); right: reconstructed trajectory using a delay embedding of one of the variables.

In Lapish et al. (2015) this technique, further backed up by machine learning methods to better segregate the flow toward different attractors (Balaguer-Ballester et al., 2011), has been used, for instance, to provide evidence for the dual-state model (Durstewitz & Seamans, 2008): Semi-attracting states reconstructed from multiple single-unit recordings during a working memory tasks in rodents (Fig. 6) were somewhat enhanced by a low dose of amphetamine ('deepening of basins of attraction'), but completely collapsed at a higher dose ('flattening of basins of attraction') which also led to significantly diminished behavioral performance. In general, of course, noise will (sometimes severely) delimit our ability to resolve the underlying system with this type of approach. We also will gain a clearer picture of only those regions of state space actually visited by the noisy experimental trajectories.

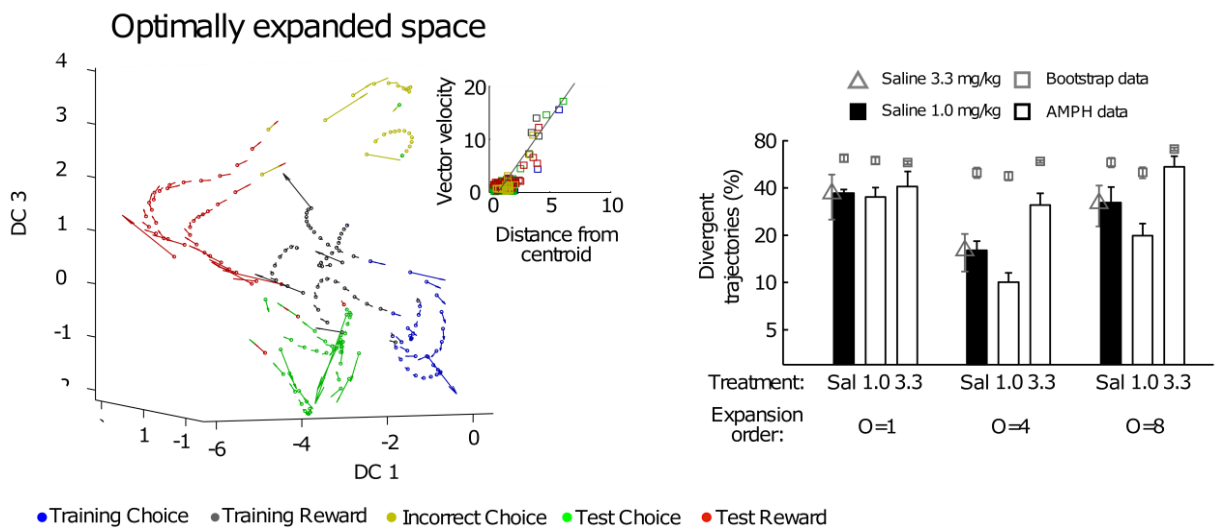

**Fig. 6.** State space reconstruction using delay embeddings and basis expansions. Left: 3D flow field projection reconstructed from multiple single-unit (MSU) recordings during a multiple-item working memory task. A combination of delay embeddings (Sauer et al., 1991) and polynomial basis expansions was used (see Balaguer-Ballester et al., 2011;

Lapish et al., 2015), and the graph was generated from the optimal (in terms of the cross-validation error) multinomial expansion order. Trajectories (velocity vectors) belonging to different task stages are color-coded. Inset: The flow slows down toward the centers of task-stage-related clusters. Right: There is, overall, convergence (attraction) to task-stage-related clusters, with only a minority of trajectories (10-20%) escaping from task-stage subspaces in the optimally expanded space (O=4). Low doses of amphetamine (1 mg/kg) enhance attraction (white center bars) while high doses (3.3 mg/kg) reduce attraction to task-stage subspaces compared to saline. Reproduced from Lapish et al. (2015).

An alternative approach is to directly infer a dynamical system's model, like the PLRNN eq. 6 (sect. 2), from the neural measurements. This type of approach implies we get access to (an approximation of) the generating system itself. Based on this we could, in principle, explore regions of state space not directly visited by the observed system dynamics, or could explicitly examine properties like the location of fixed points and their stability which are otherwise (very) hard to track directly in (highly) noisy experimental time series (i.e., through delay embeddings). In short, this approach would come with all the convenience that only an explicit mathematical formulation of the system can offer. RNNs are particularly well suited for this approach, not only since they are abstract nonlinear models of neural systems (the target of investigation in our case), but also because RNNs are dynamically universal, i.e. can emulate any dynamical system and can approximate arbitrarily closely trajectories and flow fields of any other system under certain, quite general, conditions (Funahashi & Nakamura, 1993; Kimura & Nakano, 1998; Trischler & D'Eleuterio, 2016; there are also other model systems which can do this, based for instance on polynomial basis expansions [Balaguer-Ballester et al., 2011; Brunton, Proctor, & Kutz, 2016] or reservoir computing [Lu et al., 2017], but these will not be further discussed here).

A statistical framework for inferring such process models (also called generative models) from data is known as 'state space models' (Durstewitz, Koppe, & Toutounji, 2016; Roweis & Ghahramani, 2000; Yu et al., 2006). Within this approach, we augment the PLRNN process model, eq. (6), which itself is assumed to be *unobserved (latent)*, with an observation model which links the underlying process $\{\mathbf{x}_t\}$ to the observed measurements $\{\mathbf{y}_t\}$:

(7) $p(\mathbf{y}_t|\mathbf{x}_t) = g_\lambda(f(\mathbf{x}_t))$,

where $g_\lambda$ is a probability distribution parameterized by $\lambda$, and $f$ some link function. The probabilistic mapping from RNN states $\{\mathbf{x}_t\}$ to observations $\{\mathbf{y}_t\}$ implements the usual statistical random assumptions associated with drawing a small sample from a much larger population, and with uncontrolled noise sources from the measurement process itself (e.g., from the equipment or from tissue movement). This could be simply described by a Gaussian, the most common and in the limit often adequate assumption owing to the Central Limit Theorem, but it could also take any other form (e.g., binomial or Poisson).

On top, a crucial aspect about state space models is that they assume the dynamical process itself to be noisy, which we can accommodate in eq. 6 by adding, for instance, a Gaussian noise term to the PLRNN state evolution, i.e. adding $\boldsymbol{\varepsilon}_t \sim N(\mathbf{0}, \boldsymbol{\Sigma})$ to the right hand side of eq. 6. Dynamical process noise, independent from the measurement noise, will indeed be present in most physical systems empirically observed, e.g. in the form of quantum noise, thermodynamic fluctuations, or other uncontrolled environmental factors influencing the dynamical system itself. This is particularly true for the nervous system where, e.g., the highly stochastic release of cortical synapses (Stevens, 2003) or the barrage of uncontrolled external and visceral/ somato-sensory inputs and feedbacks create profound noise sources. In fact, noise and stochasticity have been assumed to be essential ingredients of cortical computation, e.g. in decision making or for representing uncertainty in environmental outcomes (e.g., Körding & Wolpert, 2004; Pouget, Beck, Ma, & Latham, 2013), or for avoiding local minima in optimization problems the nervous system has to solve (Durstewitz, 2006). Thus, randomness in the dynamical system's temporal evolution is not only physically present, but computationally important, and both, too much dampening or inflation of cortical noise sources, may play its own role in psychiatric conditions.

Another reason for making the dynamical process itself probabilistic is to allow for some model misspecification. The fact that state space models explicitly incorporate and account for both observation and process noise sources could be another advantage over delay embedding approaches. The delay embedding theorems (Sauer et al., 1991; Takens, 1981) were originally formulated for the noise-free case only, and noise may potentially severely (Kantz & Schreiber, 2004) corrupt the reconstruction of the original attractor dynamics from a few measurements using delay embedding. State space models, on the other hand, by separating observation and process noise sources from the deterministic part of the dynamical process, could yield a clearer picture of the dynamics.

There are further objectives to a model-based approach: Like the Dynamic Causal Modeling (DCM) framework (Friston, Harrison, & Penny, 2003), which may be seen as a special instance of state space models, estimation of nonlinear dynamical systems like PLRNNs (Durstewitz, 2017b) from neural data also gives neuronally interpretable parameters. This could enable to assess, from noninvasive measurements like fMRI or EEG alone, alterations in the parameters of the underlying dynamical system in psychiatric conditions, e.g. of the synaptic connectivity $\mathbf{W}$ or the firing thresholds $\mathbf{h}$, and thus more sharply define the targets for treatment. Moreover, as noted before, it yields direct access to the stochastic process equations which most likely, within this particular class of models, generated the observed distribution of measurements, allowing for in-depth analysis (Durstewitz, 2017b).

Inferring nonlinear dynamical systems directly from experimental recordings may thus provide a powerful way to assess more directly the crucial changes in system dynamics that may underlie the observed psychiatric symptoms. At the same time it provides the kind of mathematical tool needed if the primary objective of psychiatric treatment, as argued in this article, is to reestablish a functional dynamical regime in the patient. Inferring the system dynamics this way enables to assess, track, and predict the outcome of pharmacological or behavioral treatment on exactly those properties of the brain which we claim are most central to treatment success. In this context we also reemphasize (see sect. 3) that estimating a *nonlinear* instead of a linear dynamical system is crucial, since linear systems are strongly limited in the repertoire of dynamical phenomena they can produce and hence will not capture many of the dynamical alterations proposed to underlie psychiatric diseases (see discussion in Durstewitz, 2017b).

We close this section with a brief note on how the actual statistical inference process works. As in all latent variable models, we are confronted with the problem of inferring *both* the unknown parameters *and* the unknown latent states of the underlying dynamical process, as perhaps familiar from more widely known statistical models like factor analysis, Gaussian mixture models or Hidden Markov Models (Bishop, 2006; Durstewitz, 2017a). A complication for statistical inference in these models is that for maximization of its likelihood function $p(\mathbf{Y}|\boldsymbol{\theta})$ with respect to model parameters $\boldsymbol{\theta} = \{\mathbf{A}, \mathbf{W}, \mathbf{h}, \mathbf{L}, \boldsymbol{\Sigma}, \mathbf{B}, \boldsymbol{\Gamma}\}$, or more commonly of the log-likelihood $\log p(\mathbf{Y}|\boldsymbol{\theta})$, one needs to integrate across all possible latent state paths $\mathbf{X} = \{\mathbf{x}_t\}$:

(8) $\log p(\mathbf{Y}|\boldsymbol{\theta}) = \log \int_{\mathbf{X}} p(\mathbf{Y}, \mathbf{X}|\boldsymbol{\theta}) \, d\mathbf{X}.$

This is usually not feasible through analytical derivations, and so numerical solution strategies are required. The most common one is the Expectation-Maximization (EM) algorithm (Dempster, Laird, & Rubin, 1977; McLachlan & Krishnan, 2008) which iterates state and parameter estimation until it converges to a (possibly only locally optimal) solution.

# 5.   Implications: treating dynamics

To wrap up this brief discussion of system dynamics, a central tenet in theoretical and computational neuroscience is that cognitive and emotional processes are implemented in terms of the neural dynamics (e.g. (Durstewitz, 2003; Izhikevich, 2007; Rabinovich, Huerta, Varona, et al., 2008; Wilson, 1999)). Dynamical systems theory in this sense provides the computational language of the brain. Cognitive processes can be understood in terms of trajectories traveling through neural state spaces, with the geometrical properties of these spaces, the different stable and unstable objects they harbor, their stable and unstable manifolds, nullclines, sets of homo- and heteroclinic orbits, and other features, determining their flow and thus the nature of the specific computations performed. Intrinsic systems noise, as provided e.g. by stochastic synaptic release or channel gating (Koch & Laurent, 1999; Stevens, 2003), is likely to play a fundamental role in cognitive processes, such as decision making (Pouget et al., 2013) or perceptual inference (Körding & Wolpert, 2004; Wang, 2002). Noise may push the dynamical system around among its various attractor states (causing them to be 'meta-stable'), with longer dwelling times in stronger attractors, and it may even change the dynamics itself (Longtin, 2010; for instance, in a chaotic system noise does not necessarily increase the amount of chaoticity, as one may intuitively expect, but may in fact even reduce it by stabilizing or controlling a chaotic attractor).

Computational system dynamics also provides an inherently *translational* language: We can phrase cognitive properties in animals just as in humans in the very same dynamical systems terms, in the language of state spaces, trajectories and attractors, and can thereby directly compare them. Fundamentally, the dynamical systems approach does not even care about the measurement modality, i.e. whether we reconstruct the dynamics from neuroimaging, surface electrode, multiple single-unit recordings, behavioral or self-report data. The 'only' constraints here come from the temporal and spatial resolution and the signal/noise (S/N) ratio of the recording techniques used: As long as we have any probe into the system on which all relevant degrees of freedom leave their spatio-temporal signature, and none of them is buried within the *measurement* noise (the system-intrinsic noise is part of the process), we should in principle be able to recover the same state space (guaranteed, to some degree, by the embedding theorems). Even strong spatial or temporal filtering, as is the case with fMRI for instance, may not per se be detrimental if the filtered signal amplitude still escapes the noise floor.

From this dynamical perspective, to understand how, where and why cognitive and affective processes go astray, primarily we may need to understand which crucial aspects of the stochastic system dynamics and state space geometry have changed, and perhaps not such much the biophysical and biochemical alterations underlying these changes. Vice versa, in treating psychiatric symptoms, our primary goal may be to fix the system dynamics. Assessing and monitoring the system dynamics in patients or at-risk subjects, using e.g. methods as introduced in sect. 4, may thus potentially be more informative than the typical approach of examining stable subjective phenomena by asking individuals to average over periods ranging from weeks to months, or than determining underlying biophysical or biochemical changes (of which there are often plenty with, in their complex interaction, often unknown consequences for dynamics and computation).

For instance, according to the results reported in sect. 4, the altered cognitive abilities of schizophrenic patients, and perhaps other symptoms like hallucinations as well, may ultimately be rooted in changes in prefrontal attractor dynamics, and they may interact to stabilize each other. Thus, to fix them, one would need to reestablish the healthy dynamical regime – this may be done, *but does not necessarily have to*, through dopaminergic drugs that reverse the physiological aberrations observed in schizophrenics. Indeed, if re-instantiating the healthy dynamical regime is the primary target of pharmacological treatment, a specific drug may be of little use if it restores only *part* of the ionic functions underlying the original deficits, e.g., *only* GABAergic transmission or *only* NMDA function. In fact, such drugs that reverse only some but not all of the molecular changes may make the situation even *worse*, since this could, perhaps counter-intuitively, further deteriorate functional attractor transitions because at the level of system dynamics apparently very diverse ionic effects may act *synergistically* (Durstewitz et al., 2000b). On the other hand, say we knew that changes in GABA, NMDA, and dopamine transmission were the fundamental biochemical cause of schizophrenia, then – from the dynamical systems perspective – this does not imply our drugs should target exactly these transmitter systems. Perhaps it is more convenient, easier, cheaper, or biocompatible to design and deliver compounds which target cellular calcium and potassium channels, with exactly the same implications for dynamics that restoring GABA+NMDA+dopamine function would have.

There are usually many different and partly mutually redundant routes to the same dynamical phenomena. For instance, there has been quite some discussion whether attractor multi-stability supporting working memory is, at the cellular and synaptic level, routed in recurrent NMDA synaptic drive (Brunel & Wang, 2001; Durstewitz et al., 2000a; Wang, 1999), synaptic short-term facilitation (Mongillo, Barak, & Tsodyks, 2008), $Na^+$-gated $Ca^{2+}$ channels (Egorov, Hamam, Fransén, Hasselmo, & Alonso, 2002; Fransén, Tahvildari, Egorov, Hasselmo, & Alonso, 2006), or asynchronous recurrent AMPA input (Hansel & Mato, 2001). The likely answer is that all these biophysical mechanisms contribute, perhaps to different degrees in different neural systems, and that hence one mechanism can be substituted by another if the goal is to restore functional working memory dynamics.

In our minds, this is a very important take-home: We have to think about psychiatric disorders as fundamentally rooted in aberrant network dynamics, the central layer at which biophysical and anatomical factors on the one hand side, and cognitive and emotional properties on the other, converge. If we accept this view, it may have profound implications for how we diagnose, classify, predict, and treat psychiatric symptoms. To diagnose patients, to predict or monitor drug effects and treatment outcome, we should perhaps combine functional measurements from the brain like EEG or MEG recordings with mathematical tools (sect. 4) for assessing the network dynamics from these. In designing new drugs or treatments, it would perhaps be most profitable if we exploit biophysical computer simulation models, which these days can predict physiological characteristics in quantitative detail (Hass et al., 2016; Markram et al., 2015), to explore which pharmacologically accessible parameters of the biophysical system should, and could be most easily, altered to restore functional dynamical regimes.

**Funding**

This work was supported by grants from the German Science Foundation (DFG; Du 354/8-2) and from the German Federal Ministry of Education and Research (BMBF) within the e:Med program (01ZX1311A [SP7] & 01ZX1314G [SP10]).